\documentclass[reprint,
superscriptaddress,
amsmath,amssymb,
prstper,
]{revtex4-1}

\usepackage{graphicx}
\usepackage{dcolumn}
\usepackage{bm}

\usepackage{gensymb}

\begin{document}


\title{Generation of Caustics and Spatial Rogue Waves from Nonlinear Instability}

\author{Akbar Safari}
\email[]{asafa055@uottawa.ca}
\affiliation{Department of Physics, University of Ottawa, Ottawa, ON, K1N 6N5, Canada.}

\author{Robert Fickler}
\affiliation{Department of Physics, University of Ottawa, Ottawa, ON, K1N 6N5, Canada.}

\author{Miles J. Padgett}
\affiliation{School of Physics and Astronomy, University of Glasgow, Glasgow G12 8QQ, UK.}

\author{Robert W. Boyd}
\affiliation{Department of Physics, University of Ottawa, Ottawa, ON, K1N 6N5, Canada.}
\affiliation{School of Physics and Astronomy, University of Glasgow, Glasgow G12 8QQ, UK.}
\affiliation{Institute of Optics, University of Rochester, Rochester, New York, 14627, USA.}

\vspace{3mm}

\date{\today}
%
\maketitle
\textbf{Caustics are natural phenomena in which nature concentrates the energy of waves. Although, they are known mostly in optics, caustics are intrinsic to all wave phenomena. For example, studies show that fluctuations in the profile of an ocean floor can generate random caustics and focus the energy of tsunami waves~\cite{Berry,Fleischmann}. Caustics share many similarities to rogue waves, as they both exhibit heavy-tailed distribution, i.e. an overpopulation of large events~\cite{Kharif,DudleySciRep2015}. Linear Schr\"{o}dinger-type equations are usually used to explain the wave dynamics of caustics. However, in that the wave amplitude increases dramatically in caustics, nonlinearity is inevitable in many systems. In this Letter, we investigate the effect of nonlinearity on the formation of optical caustics. We show experimentally that, in contrast to linear systems, even small phase fluctuations can generate strong caustics upon nonlinear propagation. We simulated our experiment based on the nonlinear Schr\"{o}dinger equation (NLSE) with Kerr-type nonlinearity, which describes the wave dynamics not only in optics, but also in some other physical systems such as oceans. Therefore, our results may also aid our understanding of ocean phenomena.}

Caustics can be defined as the envelope of a family of rays that define the flow of energy~\cite{Berry1980}. The energy of a wave field increases significantly on a caustic line compared to the adjacent space. When a wave acquires random phase fluctuations with correlation length larger than the wavelength of the wave, random caustics are formed upon linear propagation. Such random caustics are related to the phenomenon of branched flow observed in electron gases~\cite{Topinka} and in microwaves~\cite{Heller2010}. A familiar example of random optical caustics is the bright pattern appearing on the bottom of a swimming pool on a sunny day. Moreover, caustics are also found in large-scale wave systems such as oceans. It has been shown that a large underwater island can act as a lens and focus the energy of tsunami waves into caustics~\cite{Berry}. Subsequently, recent studies show that a small uncertainty in the profile of an ocean floor can change the caustic pattern and lead to an unexpectedly large variation in the wave height of tsunamis~\cite{Fleischmann}.

Although caustics can develop from Gaussian fluctuations, they have non-Gaussian statistics with a very long tail, meaning that waves with extremely large amplitudes appear more often than predicted from a normal distribution. The long-tailed distribution is an indication of rogue-type waves, initially studied in the context of giant waves in oceans~\cite{Kharif}. Rogue-type events are observed in various systems including optics~\cite{Solli,Dudley2014,Liu,Steinmeyer}. The formation of rogue events in (1D+1) waves (one spatial dimension plus time) has been extensively studied. Nonlinearity, namely modulational instability, is commonly used to explain how rogue waves develop in these (1D+1) systems such as unidirectional water waves and optical fibers. However, the dynamics of waves is richer in higher dimensions where rogue waves can be formed from spatial focusing of waves without the aid of nonlinearity. In fact, numerous studies have shown that concentration of waves in caustics is a linear mechanism that can generate spatial rogue waves in oceans~\cite{Brown, Fochesato, Heller2008, Peregrine1983} and also in optics~\cite{DudleySciRep2015}. Moreover, the role of nonlinearity in the formation of rogue waves is still under debate. In optical systems, nonlinearity can either trigger~\cite{DelRe} or destroy~\cite{Mattheakis} rogue wave events. Similarly, recent studies explain oceanic rogue waves without modulational instability~\cite{DudleySciRep2016} or any type of nonlinearity~\cite{Steinmeyer2016}.

In this Letter, we investigate the effect of nonlinearity on the formation of optical caustics in (2D+1), where light propagates in two transverse directions plus one longitudinal direction along the beam axis, $z$. This effect has been studied previously in the context of nonlinear wave-current interactions in oceans~\cite{Peregrine1983,Peregrine1979}, and it has been stated that nonlinearity may wash out caustics and decreases the amplitude of extreme waves by destroying the phase coherence~\cite{Kharif}. However, to our knowledge, the effect of nonlinear instability on the formation of caustics is not well examined. We show that in contrast to linear propagation where relatively large fluctuations are required~\cite{DudleySciRep2015}, even small phase fluctuations can generate sharp caustics with the aid of nonlinear instability in the spatial propagation.

\begin{figure}                    
\begin{center}
\includegraphics[width=8.5cm]{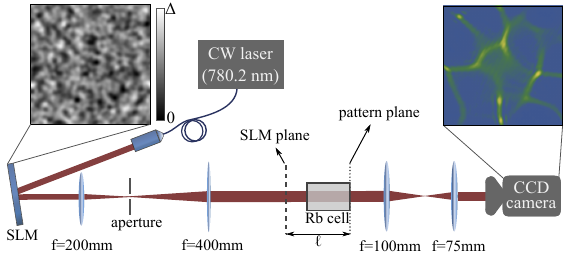} 
\end{center}
\caption{\textbf{Scheme of the experimental setup.} The spatial light modulator (SLM) imprints a random phase mask (upper-left inset) onto the transverse profile of a cw laser beam. The first imaging system images the SLM onto the plane shown by the dashed line (SLM plane). At this point the transverse intensity distribution of the beam follows the Gaussian profile of the input laser. An intensity pattern develops upon propagation. The distance $l$, after which the sharpest pattern is formed depends on the amplitude $\Delta$ of the phase modulation. Another imaging system is used to image the pattern plane (dotted line) onto the CCD camera. The upper-right inset shows an example pattern generated from $\Delta=8\pi$. To study nonlinear propagation, the Rb cell is placed in the end of the propagation before the pattern plane.}
\label{Setup} 
\end{figure}

We first study caustic formation for the case of linear propagation through free space. In order to generate optical caustics in the laboratory, we use collimated continuous wave (cw) laser light with a beam waist of $w_0\simeq 1$ mm, and modulate its phase front with a smooth random phase mask. We implement this random phase modulation by forming a computed generated hologram on a spatial light modulator (SLM) to create a phase mask. The hologram is blazed to maximize the efficiency of the first diffracted order, which is separated from the other orders by use of an aperture. The random phase across the mask has a Gaussian distribution with correlation length $\delta=150$~$\mu$m and an amplitude $\Delta$ that can vary up to $16\pi$ (Fig.~\ref{Setup}). An imaging system is used to image the SLM plane and expand the beam by a factor of two. Upon propagation in free space, a pattern is formed that is imaged onto a CCD camera (640$\times$640 pixels and 8-bit depth) with another imaging system. The recorded structure of the pattern depends on the random phase mask displayed on the SLM. The amplitude of the imprinted phase determines the strength of the intensity maxima in the caustics and the distance $l$ at which the sharpest pattern is formed. The degree of sharpness can be quantified by the scintillation index defined by~\cite{Stockmann,Kravtsov}
\begin{eqnarray}                       
\beta ^2 = \frac{\langle I^2 \rangle - \langle I \rangle ^2}{\langle I \rangle ^2},
\label{Scint}
\end{eqnarray}
where $\langle ... \rangle$ indicates the spatial average over the transverse plane. Speckle patterns that obey Gaussian statistics, for example, have a scintillation index of unity. A scintillation index above unity indicates the strength of concentration of light with respect to the adjacent space. Thus, the sharper the caustic, the higher the scintillation index. For our system, we found out that when $\Delta$ is greater than $6\pi$, the scintillation index goes above unity after 7.5 cm (the length of our nonlinear medium employed later) of propagation in linear space. Therefore, we take this value as the caustic threshold in our experiment. One should note that this threshold is not universal and depends on parameters such as the wavelength and the correlation length, $\delta$.

\begin{figure}                   
\includegraphics[width=8.5cm]{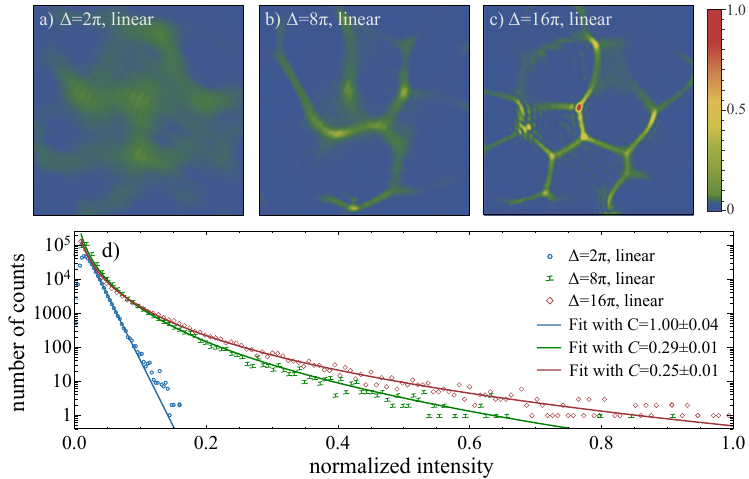}
\caption{\textbf{Generation of caustics upon linear propagation.} a, b, c) Examples of patterns formed after propagation in free space from different phase masks with amplitudes $\Delta=2\pi$, $8\pi$ and $16\pi$, respectively. We see that sharp caustics are formed only under strong phase modulation. d) Intensity distributions for the patterns generated upon propagation through free space from three different phase amplitudes $\Delta$. The $C$-parameter from the fit function characterizes the heavy-tailed behaviour; the lower the $C$-parameter, the longer the tail of the distribution. Thus, sharp caustics are distinguished by their heavy-tailed statistics.} 
\label{LinCaus} 
\end{figure}

Figure~\ref{LinCaus}(a), (b) and (c) show the sharpest patterns formed from three different phase masks with amplitudes $\Delta=2\pi$, $8\pi$ and $16\pi$, after propagation distances of $l= 27.5$ cm, $10.5$ cm and $7.5$ cm, respectively. This figure clearly shows that a sharp caustic is formed only if the phase variations are large. When the phase variations are small compared to the caustic threshold, $6\pi$, the power of the caustic focus is weak, the pattern is blurred and is formed at a longer distance. To verify that this observation holds in general and not only for the patterns shown, we generated a set of 1000 uncorrelated phase masks for each amplitude $\Delta$, and evaluated the statistical distribution of intensities in the final patterns. Since the Gaussian transverse profile of the laser beam is superimposed on the patterns, one does not obtain an unbiased intensity distribution within each pattern. Therefore, we considered the intensities of pixels along a circular annulus centred on the beam axis. As shown in Fig.~\ref{LinCaus}(d) (blue circles) when the phase modulation is weak ($\Delta = 2\pi$) the statistical distribution closely follows an exponential decay. However, as mentioned earlier, caustic patterns have heavy-tailed statistics~\cite{Mercier,Nye}. This is confirmed in our measurements, Fig.~\ref{LinCaus}(d); as the phase variations get stronger, so do the intensity fluctuations and the statistics become non-Gaussian. One should note that the histograms are plotted in logarithmic scale and all intensities in Fig.~\ref{LinCaus} are normalized to the maximum intensity observed with $\Delta=16\pi$. 

In order to quantify the heavy-tailed behaviour we have used a least-square method to fit the distribution with a stretched exponential function $A$ $\exp(-B I^C)$, where $A$, $B$ and $C$ are the fitting parameters and $I$ is the normalized intensity~\cite{DelRe}. We are interested primarily in the $C$-parameter as it indicates the curvature of the function and quantifies the heavy-tailed behaviour. In speckle patterns generated from random scatterer, the intensity obeys negative exponential statistics, and thus $C=1$. As $C$ becomes smaller, the tail of the distribution gets longer. The solid lines in Fig~\ref{LinCaus}(d) show the fit functions with the indicated $C$-parameters. As expected, the patterns generated from strong phase modulations ($\Delta = 16 \pi$) have the lowest $C$-parameters, and thus the strongest rogue-type statistics.

\begin{figure}                   
\includegraphics[width=8.5cm]{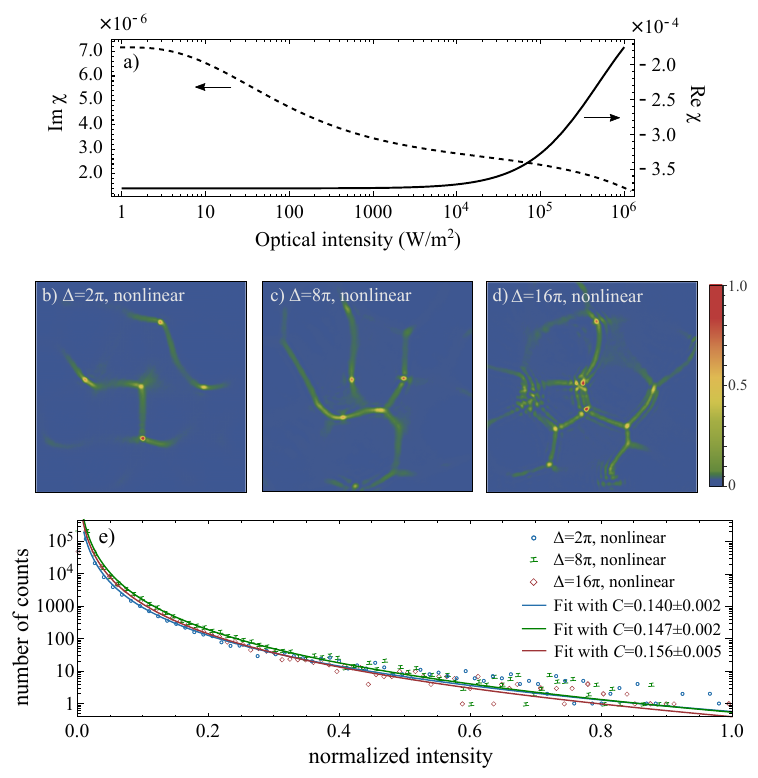}
\caption{\textbf{Generation of caustics upon nonlinear propagation.} a) Real and imaginary parts of the total susceptibility of Rb vapour at a temperature of 115 $^\circ$C and the frequency detuning used in the experiment (see the text). Re $\chi$ and thus the refractive index increases with intensity, indicating nonlinear focusing. b, c, d) Caustic patterns generated from the same phase masks as in Fig.~\ref{LinCaus}, but after the nonlinear propagation in Rb. In contrast to the linear case shown in Fig.~\ref{LinCaus}, even small phase modulations, with the aid of nonlinear focusing, can concentrate light into sharp caustics. e) Intensity distributions of the nonlinear caustic patterns generated from three different phase amplitudes $\Delta$. In this case, all the patterns have approximately the same statistics, irrespective of the magnitude of $\Delta$.} 
\label{NLCaus} 
\end{figure}

To investigate the effect of nonlinearity on the formation of caustics we use rubidium (Rb) vapour as the nonlinear medium. The motivation for using atomic vapours as the nonlinear medium is that they can be saturated easily and thus show large nonlinearity that can be controlled simply through the laser frequency detuning. The Rb cell is 7.5 cm long and is filled with natural Rb: $^{85}$Rb and $^{87}$Rb with abundances of $72\%$ and $28\%$, respectively. The Rb cell is heated to 115 $^\circ$C and the laser is blue detuned by 840 MHz from the $5^2S_{1/2}$, $F=3$ $\rightarrow$ $5^2P_{3/2}$, $F=4$ transition in $^{85}$Rb. The cell is placed in the setup (Fig.~\ref{Setup}) such that the last 7.5 cm of propagation before the caustic pattern is formed in the linear case is now taking place within Rb. The camera images the output of the cell. The laser power at the entrance of the cell is approximately 140 mW. Real (Re) and imaginary (Im) parts of the total susceptibility are calculated from our theoretical model based on ref.~\cite{BoydNL}. Doppler broadening is implemented by calculating the convolution of the power broadened lineshape with the Gaussian distribution of the atomic velocities~\cite{Loudon}. An effective saturation intensity is incorporated to take into account the effect of optical pumping~\cite{Daniel}. As shown in Fig.~\ref{NLCaus}(a), Re $\chi$ and consequently the refractive index $n\simeq 1+$Re $\chi$/2 increases with intensity. Thus, self-focusing is expected at this frequency detuning. The maximum nonlinear phase shift experienced by the laser light in passing through the Rb cell is approximately 4$\pi$ radians. Im $\chi$ and therefore absorption decreases with intensity, indicating saturation of absorption. 

For direct demonstration of the effect of nonlinearity, we employed the same sequence of phase masks as for the study of linear propagation. The resulting patterns after nonlinear propagation through the Rb cell are shown in Fig~\ref{NLCaus}. A comparison of these patterns with those of Fig.~\ref{LinCaus} for linear propagation indicates that nonlinear instability in spatial propagation enhances the sharpness of the patterns without changing their overall structure. This enhancement is more profound when the linear caustic is weak, i.e. when the phase modulation is not strong enough to form sharp caustics for linear propagation. The intensity distribution shown in Fig.~\ref{NLCaus}(e) confirms this observation statistically. When the phase modulation is weak ($\Delta=2\pi$) nonlinearity changes the statistical distribution of intensities substantially. Conversely, the distribution is less affected by nonlinearity for $\Delta=16\pi$. This is in contrast to the conclusions reached in ref.~\cite{Mattheakis} where the heavy-tailed distribution is found to be suppressed by nonlinear propagation. In our experiment, in the presence of nonlinearity, all patterns have approximately the same heavy-tailed statistics and the same maximum intensity, irrespective of the amplitudes of the phase modulations. We have normalized all intensities in Fig.~\ref{NLCaus} to this maximum intensity, which is about 25\% of the maximum intensity in Fig.~\ref{LinCaus}. Although this decrease in intensity, which is due to linear absorption, reduces the strength of nonlinearity, it does not play a central role in our results. Moreover, absorption can be neglected when the laser frequency is far from resonance and a longer nonlinear medium is used~\citep{McCormick}.

\begin{figure}                   
\begin{center}
\includegraphics[width=8.5cm]{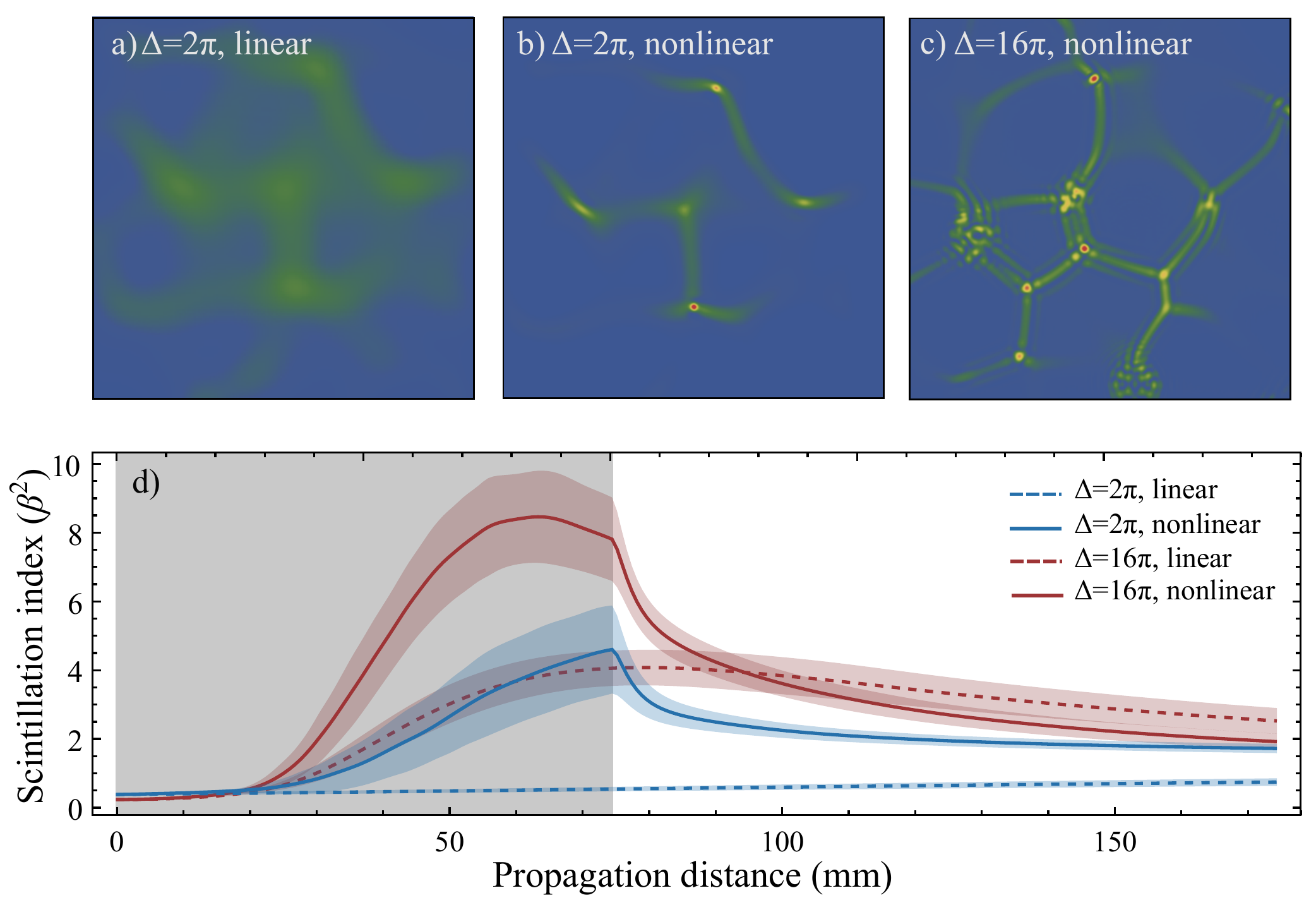} 
\end{center}
\caption{\textbf{Intensity patterns and scintillation indices from computer simulation.} a, b, c) Patterns obtained from the numerical simulation, which show excellent agreement with the experimental results shown in Figs.~\ref{LinCaus} and \ref{NLCaus}. We have normalized the intensities in the same way as in the experimental images, as explained in the text. d) Scintillation indices, $\beta^2$, averaged over 1000 patterns, calculated from our numerical simulation, as a function of the propagation distance from the entrance of the Rb cell up to 100 mm after the cell where partial speckles are formed. The grey area indicates the 7.5-cm-long Rb cell for the nonlinear cases and the shaded regions show one standard deviation from the mean. The nonlinear patterns have larger scintillation indices compared to the corresponding linear cases. A scintillation index greater than unity indicates the presence of sharp caustics.} 
\label{Simul} 
\end{figure}

For further investigations and to test the role of Kerr nonlinearity on the formation of nonlinear caustics, we simulated the results of our experiment numerically using a beam propagation method based on the use of the fast Fourier transform. Similar to the propagation of waves in fluids, the propagation of the laser field through Rb vapour is described by the nonlinear Schr\"{o}dinger equation (NLSE)
\begin{eqnarray}                       
\frac{\partial \mathcal{E}}{\partial z} - \frac{i}{2k} \nabla _\bot ^2 \mathcal{E} = \frac{ik}{2\epsilon_0} P,
\label{NLSE} 
\end{eqnarray}
where $\mathcal{E}$ is the field amplitude defined by $E=\mathcal{E}e^{i(kz-\omega t)}+$c.c. and $\nabla _\bot ^2=\partial^2 / \partial x^2 + \partial^2 / \partial y^2$ is the transverse Laplacian. For a purely third-order nonlinear medium the atomic polarization is given by $P=3 \epsilon_0 \chi^{(3)}|\mathcal{E}|^2 \mathcal{E}$. However, to include higher order effects, we use the more general form $P=\epsilon_0 \chi \mathcal{E}$. The total susceptibility, $\chi$, is taken from our Rb numerical model, as shown in Fig.~\ref{NLCaus}(a), without any adjustable parameters. The split-step method~\cite{Agrawal} was used to implement nonlinearity in the simulation. To be able to compare the results directly with the experimental patterns, we used the same random phase masks as in the experiment. All simulation results for both linear and nonlinear cases are in extremely good agreement with the experiment (example patterns from the simulation are shown in Fig.~\ref{Simul}). This excellent matching of the results confirms that Kerr-saturated nonlinearity is the mechanism behind the generation of caustics from small phase fluctuations.

Since experimental imaging techniques are not compatible with nonlinear propagation, we cannot use imaging to determine the patterns within the Rb cell. However, our numerical simulation reproduces the experimental results accurately. We are thus confident in using our numerical method to study the patterns within the nonlinear medium. We use the scintillation index (Eq.~\ref{Scint}) to characterize the sharpness of the caustics. Fig.~\ref{Simul}(d) shows how the scintillation indices vary in linear and nonlinear propagation from the entrance of the Rb cell up to 100 mm after the cell, where partial speckles are formed. Inside the Rb cell, the nonlinear focusing exceeds diffraction and thus accentuates the caustic focusing. Therefore, the scintillation index tends to be large and to increase with propagation distance. After the cell, the scintillation index drops very quickly as the result of diffraction.

In conclusion, linear caustics and nonlinear instability are known to be responsible for focusing the energy of waves and for generating rogue-type events in various systems. Here, we experimentally and numerically investigated wave dynamics in the presence of both mechanisms. Our results show that the formation of caustics in Kerr media requires significantly smaller fluctuations compared to linear propagation. Thus, nonlinear instability in spatial propagation amplifies even small phase fluctuations and generates rogue-type waves with very large amplitudes. The effect of nonlinear focusing on the statistics is more important when the phase modulation is weak. In contrast, when the linear caustic is sufficiently sharp, nonlinearity does not change the distribution significantly. Our experiment was performed in a nonlinear optical system, and the NLSE was used to simulate the dynamics. Importantly, the NLSE also describes the nonlinear wave propagation in different physical systems, such as fluids~\cite{Whitham} and Bose-Einstein condensates~\cite{Pitaevskii}, both of which exhibit caustics as well~\cite{Bludov}. Therefore, the nonlinear generation and enhancement of caustics and rogue-type events, which we observed, are not limited to optics and might also be realized in other physical systems.

\subsection{Acknowledgement}
This work was supported by the Canada Excellence Research Chairs program and the National Science and Engineering Research Council of Canada (NSERC). R.F. also acknowledges the support of the Banting postdoctoral fellowship of NSERC.

\bibliography{Ref}

\begin{thebibliography}{32}%
\makeatletter
\providecommand \@ifxundefined [1]{%
 \@ifx{#1\undefined}
}%
\providecommand \@ifnum [1]{%
 \ifnum #1\expandafter \@firstoftwo
 \else \expandafter \@secondoftwo
 \fi
}%
\providecommand \@ifx [1]{%
 \ifx #1\expandafter \@firstoftwo
 \else \expandafter \@secondoftwo
 \fi
}%
\providecommand \natexlab [1]{#1}%
\providecommand \enquote  [1]{``#1''}%
\providecommand \bibnamefont  [1]{#1}%
\providecommand \bibfnamefont [1]{#1}%
\providecommand \citenamefont [1]{#1}%
\providecommand \href@noop [0]{\@secondoftwo}%
\providecommand \href [0]{\begingroup \@sanitize@url \@href}%
\providecommand \@href[1]{\@@startlink{#1}\@@href}%
\providecommand \@@href[1]{\endgroup#1\@@endlink}%
\providecommand \@sanitize@url [0]{\catcode `\\12\catcode `\$12\catcode
  `\&12\catcode `\#12\catcode `\^12\catcode `\_12\catcode `\%12\relax}%
\providecommand \@@startlink[1]{}%
\providecommand \@@endlink[0]{}%
\providecommand \url  [0]{\begingroup\@sanitize@url \@url }%
\providecommand \@url [1]{\endgroup\@href {#1}{\urlprefix }}%
\providecommand \urlprefix  [0]{URL }%
\providecommand \Eprint [0]{\href }%
\providecommand \doibase [0]{http://dx.doi.org/}%
\providecommand \selectlanguage [0]{\@gobble}%
\providecommand \bibinfo  [0]{\@secondoftwo}%
\providecommand \bibfield  [0]{\@secondoftwo}%
\providecommand \translation [1]{[#1]}%
\providecommand \BibitemOpen [0]{}%
\providecommand \bibitemStop [0]{}%
\providecommand \bibitemNoStop [0]{.\EOS\space}%
\providecommand \EOS [0]{\spacefactor3000\relax}%
\providecommand \BibitemShut  [1]{\csname bibitem#1\endcsname}%
\let\auto@bib@innerbib\@empty
\bibitem [{\citenamefont {Berry}(2007)}]{Berry}%
  \BibitemOpen
  \bibfield  {author} {\bibinfo {author} {\bibfnamefont {M.V}\ \bibnamefont
  {Berry}},\ }\bibfield  {title} {\enquote {\bibinfo {title} {Focused tsunami
  waves},}\ }\href {\doibase 10.1098/rspa.2007.0051} {\bibfield  {journal}
  {\bibinfo  {journal} {Proceedings of the Royal Society of London A:
  Mathematical, Physical and Engineering Sciences}\ }\textbf {\bibinfo {volume}
  {463}},\ \bibinfo {pages} {3055--3071} (\bibinfo {year} {2007})}\BibitemShut
  {NoStop}%
\bibitem [{\citenamefont {Degueldre}\ \emph {et~al.}(2016)\citenamefont
  {Degueldre}, \citenamefont {Metzger}, \citenamefont {Geisel},\ and\
  \citenamefont {Fleischmann}}]{Fleischmann}%
  \BibitemOpen
  \bibfield  {author} {\bibinfo {author} {\bibfnamefont {Henri}\ \bibnamefont
  {Degueldre}}, \bibinfo {author} {\bibfnamefont {Jakob~J.}\ \bibnamefont
  {Metzger}}, \bibinfo {author} {\bibfnamefont {Theo}\ \bibnamefont {Geisel}},
  \ and\ \bibinfo {author} {\bibfnamefont {Ragnar}\ \bibnamefont
  {Fleischmann}},\ }\bibfield  {title} {\enquote {\bibinfo {title} {Random
  focusing of tsunami waves},}\ }\href {http://dx.doi.org/10.1038/nphys3557}
  {\bibfield  {journal} {\bibinfo  {journal} {Nat Phys}\ }\textbf {\bibinfo
  {volume} {12}},\ \bibinfo {pages} {259} (\bibinfo {year} {2016})}\BibitemShut
  {NoStop}%
\bibitem [{\citenamefont {Kharif}\ and\ \citenamefont
  {Pelinovsky}(2003)}]{Kharif}%
  \BibitemOpen
  \bibfield  {author} {\bibinfo {author} {\bibfnamefont {Christian}\
  \bibnamefont {Kharif}}\ and\ \bibinfo {author} {\bibfnamefont {Efim}\
  \bibnamefont {Pelinovsky}},\ }\bibfield  {title} {\enquote {\bibinfo {title}
  {Physical mechanisms of the rogue wave phenomenon},}\ }\href {\doibase
  http://dx.doi.org/10.1016/j.euromechflu.2003.09.002} {\bibfield  {journal}
  {\bibinfo  {journal} {European Journal of Mechanics - B/Fluids}\ }\textbf
  {\bibinfo {volume} {22}},\ \bibinfo {pages} {603 -- 634} (\bibinfo {year}
  {2003})}\BibitemShut {NoStop}%
\bibitem [{\citenamefont {Mathis}\ \emph {et~al.}(2015)\citenamefont {Mathis},
  \citenamefont {Froehly}, \citenamefont {Toenger}, \citenamefont {Dias},
  \citenamefont {Genty},\ and\ \citenamefont {Dudley}}]{DudleySciRep2015}%
  \BibitemOpen
  \bibfield  {author} {\bibinfo {author} {\bibfnamefont {Amaury}\ \bibnamefont
  {Mathis}}, \bibinfo {author} {\bibfnamefont {Luc}\ \bibnamefont {Froehly}},
  \bibinfo {author} {\bibfnamefont {Shanti}\ \bibnamefont {Toenger}}, \bibinfo
  {author} {\bibfnamefont {Fr\'ed\'eric}\ \bibnamefont {Dias}}, \bibinfo
  {author} {\bibfnamefont {Goery}\ \bibnamefont {Genty}}, \ and\ \bibinfo
  {author} {\bibfnamefont {John~M.}\ \bibnamefont {Dudley}},\ }\bibfield
  {title} {\enquote {\bibinfo {title} {Caustics and rogue waves in an optical
  sea},}\ }\href {\doibase 10.1038/srep12822} {\bibfield  {journal} {\bibinfo
  {journal} {Scientific Reports}\ }\textbf {\bibinfo {volume} {5}},\ \bibinfo
  {pages} {12822} (\bibinfo {year} {2015})}\BibitemShut {NoStop}%
\bibitem [{\citenamefont {Berry}\ and\ \citenamefont
  {Upstill}(1980)}]{Berry1980}%
  \BibitemOpen
  \bibfield  {author} {\bibinfo {author} {\bibfnamefont {M.V.}\ \bibnamefont
  {Berry}}\ and\ \bibinfo {author} {\bibfnamefont {C.}~\bibnamefont
  {Upstill}},\ }\bibfield  {title} {\enquote {\bibinfo {title} {Iv catastrophe
  optics: Morphologies of caustics and their diffraction patterns},}\ }\href
  {\doibase http://dx.doi.org/10.1016/S0079-6638(08)70215-4} {\bibfield
  {journal} {\bibinfo  {journal} {Progress in Optics}\ }\textbf {\bibinfo
  {volume} {18}},\ \bibinfo {pages} {257 -- 346} (\bibinfo {year}
  {1980})}\BibitemShut {NoStop}%
\bibitem [{\citenamefont {Topinka}\ \emph {et~al.}(2001)\citenamefont
  {Topinka}, \citenamefont {LeRoy}, \citenamefont {Westervelt}, \citenamefont
  {Shaw}, \citenamefont {Fleischmann}, \citenamefont {Heller}, \citenamefont
  {Maranowski},\ and\ \citenamefont {Gossard}}]{Topinka}%
  \BibitemOpen
  \bibfield  {author} {\bibinfo {author} {\bibfnamefont {M.~A.}\ \bibnamefont
  {Topinka}}, \bibinfo {author} {\bibfnamefont {B.~J.}\ \bibnamefont {LeRoy}},
  \bibinfo {author} {\bibfnamefont {R.~M.}\ \bibnamefont {Westervelt}},
  \bibinfo {author} {\bibfnamefont {S.~E.~J.}\ \bibnamefont {Shaw}}, \bibinfo
  {author} {\bibfnamefont {R.}~\bibnamefont {Fleischmann}}, \bibinfo {author}
  {\bibfnamefont {E.~J.}\ \bibnamefont {Heller}}, \bibinfo {author}
  {\bibfnamefont {K.~D.}\ \bibnamefont {Maranowski}}, \ and\ \bibinfo {author}
  {\bibfnamefont {A.~C.}\ \bibnamefont {Gossard}},\ }\bibfield  {title}
  {\enquote {\bibinfo {title} {Coherent branched flow in a two-dimensional
  electron gas},}\ }\href {http://dx.doi.org/10.1038/35065553} {\bibfield
  {journal} {\bibinfo  {journal} {Nature}\ }\textbf {\bibinfo {volume} {410}},\
  \bibinfo {pages} {183} (\bibinfo {year} {2001})}\BibitemShut {NoStop}%
\bibitem [{\citenamefont {H\"ohmann}\ \emph {et~al.}(2010)\citenamefont
  {H\"ohmann}, \citenamefont {Kuhl}, \citenamefont {St\"ockmann}, \citenamefont
  {Kaplan},\ and\ \citenamefont {Heller}}]{Heller2010}%
  \BibitemOpen
  \bibfield  {author} {\bibinfo {author} {\bibfnamefont {R.}~\bibnamefont
  {H\"ohmann}}, \bibinfo {author} {\bibfnamefont {U.}~\bibnamefont {Kuhl}},
  \bibinfo {author} {\bibfnamefont {H.-J.}\ \bibnamefont {St\"ockmann}},
  \bibinfo {author} {\bibfnamefont {L.}~\bibnamefont {Kaplan}}, \ and\ \bibinfo
  {author} {\bibfnamefont {E.~J.}\ \bibnamefont {Heller}},\ }\bibfield  {title}
  {\enquote {\bibinfo {title} {Freak waves in the linear regime: A microwave
  study},}\ }\href {\doibase 10.1103/PhysRevLett.104.093901} {\bibfield
  {journal} {\bibinfo  {journal} {Phys. Rev. Lett.}\ }\textbf {\bibinfo
  {volume} {104}},\ \bibinfo {pages} {093901} (\bibinfo {year}
  {2010})}\BibitemShut {NoStop}%
\bibitem [{\citenamefont {Solli}\ \emph {et~al.}(2007)\citenamefont {Solli},
  \citenamefont {Ropers}, \citenamefont {Koonath},\ and\ \citenamefont
  {Jalali}}]{Solli}%
  \BibitemOpen
  \bibfield  {author} {\bibinfo {author} {\bibfnamefont {D.~R.}\ \bibnamefont
  {Solli}}, \bibinfo {author} {\bibfnamefont {C.}~\bibnamefont {Ropers}},
  \bibinfo {author} {\bibfnamefont {P.}~\bibnamefont {Koonath}}, \ and\
  \bibinfo {author} {\bibfnamefont {B.}~\bibnamefont {Jalali}},\ }\bibfield
  {title} {\enquote {\bibinfo {title} {Optical rogue waves},}\ }\href {\doibase
  10.1038/nature06402} {\bibfield  {journal} {\bibinfo  {journal} {Nature}\
  }\textbf {\bibinfo {volume} {450}},\ \bibinfo {pages} {1054--1057} (\bibinfo
  {year} {2007})}\BibitemShut {NoStop}%
\bibitem [{\citenamefont {Dudley}\ \emph {et~al.}(2014)\citenamefont {Dudley},
  \citenamefont {Dias}, \citenamefont {Erkintalo},\ and\ \citenamefont
  {Genty}}]{Dudley2014}%
  \BibitemOpen
  \bibfield  {author} {\bibinfo {author} {\bibfnamefont {John~M.}\ \bibnamefont
  {Dudley}}, \bibinfo {author} {\bibfnamefont {Fr\'ed\'eric}\ \bibnamefont
  {Dias}}, \bibinfo {author} {\bibfnamefont {Miro}\ \bibnamefont {Erkintalo}},
  \ and\ \bibinfo {author} {\bibfnamefont {Goery}\ \bibnamefont {Genty}},\
  }\bibfield  {title} {\enquote {\bibinfo {title} {Instabilities, breathers and
  rogue waves in optics},}\ }\href {\doibase 10.1038/nphoton.2014.220}
  {\bibfield  {journal} {\bibinfo  {journal} {Nature Photonics}\ }\textbf
  {\bibinfo {volume} {8}},\ \bibinfo {pages} {755–764} (\bibinfo {year}
  {2014})}\BibitemShut {NoStop}%
\bibitem [{\citenamefont {Liu}\ \emph {et~al.}(2015)\citenamefont {Liu},
  \citenamefont {van~der Wel}, \citenamefont {Rotenberg}, \citenamefont
  {Kuipers}, \citenamefont {Krauss}, \citenamefont {Falco},\ and\ \citenamefont
  {Fratalocchi}}]{Liu}%
  \BibitemOpen
  \bibfield  {author} {\bibinfo {author} {\bibfnamefont {C.}~\bibnamefont
  {Liu}}, \bibinfo {author} {\bibfnamefont {R.~E.~C.}\ \bibnamefont {van~der
  Wel}}, \bibinfo {author} {\bibfnamefont {N.}~\bibnamefont {Rotenberg}},
  \bibinfo {author} {\bibfnamefont {L.}~\bibnamefont {Kuipers}}, \bibinfo
  {author} {\bibfnamefont {T.~F.}\ \bibnamefont {Krauss}}, \bibinfo {author}
  {\bibfnamefont {A.~Di}\ \bibnamefont {Falco}}, \ and\ \bibinfo {author}
  {\bibfnamefont {A.}~\bibnamefont {Fratalocchi}},\ }\bibfield  {title}
  {\enquote {\bibinfo {title} {Triggering extreme events at the nanoscale in
  photonic seas},}\ }\href {\doibase 10.1038/nphys3263} {\bibfield  {journal}
  {\bibinfo  {journal} {Nature Physics}\ }\textbf {\bibinfo {volume} {11}},\
  \bibinfo {pages} {358–363} (\bibinfo {year} {2015})}\BibitemShut {NoStop}%
\bibitem [{\citenamefont {Birkholz}\ \emph {et~al.}(2013)\citenamefont
  {Birkholz}, \citenamefont {Nibbering}, \citenamefont {Br\'ee}, \citenamefont
  {Skupin}, \citenamefont {Demircan}, \citenamefont {Genty},\ and\
  \citenamefont {Steinmeyer}}]{Steinmeyer}%
  \BibitemOpen
  \bibfield  {author} {\bibinfo {author} {\bibfnamefont {Simon}\ \bibnamefont
  {Birkholz}}, \bibinfo {author} {\bibfnamefont {Erik T.~J.}\ \bibnamefont
  {Nibbering}}, \bibinfo {author} {\bibfnamefont {Carsten}\ \bibnamefont
  {Br\'ee}}, \bibinfo {author} {\bibfnamefont {Stefan}\ \bibnamefont {Skupin}},
  \bibinfo {author} {\bibfnamefont {Ayhan}\ \bibnamefont {Demircan}}, \bibinfo
  {author} {\bibfnamefont {Go\"ery}\ \bibnamefont {Genty}}, \ and\ \bibinfo
  {author} {\bibfnamefont {G\"unter}\ \bibnamefont {Steinmeyer}},\ }\bibfield
  {title} {\enquote {\bibinfo {title} {Spatiotemporal rogue events in optical
  multiple filamentation},}\ }\href {\doibase 10.1103/PhysRevLett.111.243903}
  {\bibfield  {journal} {\bibinfo  {journal} {Phys. Rev. Lett.}\ }\textbf
  {\bibinfo {volume} {111}},\ \bibinfo {pages} {243903} (\bibinfo {year}
  {2013})}\BibitemShut {NoStop}%
\bibitem [{\citenamefont {Brown}(2001)}]{Brown}%
  \BibitemOpen
  \bibfield  {author} {\bibinfo {author} {\bibfnamefont {Michael~G.}\
  \bibnamefont {Brown}},\ }\bibfield  {title} {\enquote {\bibinfo {title}
  {Space–time surface gravity wave caustics: structurally stable extreme wave
  events},}\ }\href {\doibase http://dx.doi.org/10.1016/S0165-2125(00)00054-8}
  {\bibfield  {journal} {\bibinfo  {journal} {Wave Motion}\ }\textbf {\bibinfo
  {volume} {33}},\ \bibinfo {pages} {117 -- 143} (\bibinfo {year}
  {2001})}\BibitemShut {NoStop}%
\bibitem [{\citenamefont {Fochesato}\ \emph {et~al.}(2007)\citenamefont
  {Fochesato}, \citenamefont {Grilli},\ and\ \citenamefont {Dias}}]{Fochesato}%
  \BibitemOpen
  \bibfield  {author} {\bibinfo {author} {\bibfnamefont {Christophe}\
  \bibnamefont {Fochesato}}, \bibinfo {author} {\bibfnamefont {St\'ephan}\
  \bibnamefont {Grilli}}, \ and\ \bibinfo {author} {\bibfnamefont
  {Fr\'ed\'eric}\ \bibnamefont {Dias}},\ }\bibfield  {title} {\enquote
  {\bibinfo {title} {Numerical modeling of extreme rogue waves generated by
  directional energy focusing},}\ }\href {\doibase
  http://dx.doi.org/10.1016/j.wavemoti.2007.01.003} {\bibfield  {journal}
  {\bibinfo  {journal} {Wave Motion}\ }\textbf {\bibinfo {volume} {44}},\
  \bibinfo {pages} {395 -- 416} (\bibinfo {year} {2007})}\BibitemShut {NoStop}%
\bibitem [{\citenamefont {Heller}\ \emph {et~al.}(2008)\citenamefont {Heller},
  \citenamefont {Kaplan},\ and\ \citenamefont {Dahlen}}]{Heller2008}%
  \BibitemOpen
  \bibfield  {author} {\bibinfo {author} {\bibfnamefont {E.~J.}\ \bibnamefont
  {Heller}}, \bibinfo {author} {\bibfnamefont {L.}~\bibnamefont {Kaplan}}, \
  and\ \bibinfo {author} {\bibfnamefont {A.}~\bibnamefont {Dahlen}},\
  }\bibfield  {title} {\enquote {\bibinfo {title} {Refraction of a gaussian
  seaway},}\ }\href {\doibase 10.1029/2008JC004748} {\bibfield  {journal}
  {\bibinfo  {journal} {J. Geophys. Res.}\ }\textbf {\bibinfo {volume} {113}},\
  \bibinfo {pages} {C09023} (\bibinfo {year} {2008})}\BibitemShut {NoStop}%
\bibitem [{\citenamefont {Peregrine}(1983)}]{Peregrine1983}%
  \BibitemOpen
  \bibfield  {author} {\bibinfo {author} {\bibfnamefont {D.H.}\ \bibnamefont
  {Peregrine}},\ }\bibfield  {title} {\enquote {\bibinfo {title} {Wave jumps
  and caustics in the propagation of finite-amplitude water waves},}\ }\href
  {\doibase 10.1017/S0022112083002220} {\bibfield  {journal} {\bibinfo
  {journal} {J. Fluid Mech}\ }\textbf {\bibinfo {volume} {136}},\ \bibinfo
  {pages} {435} (\bibinfo {year} {1983})}\BibitemShut {NoStop}%
\bibitem [{\citenamefont {Pierangeli}\ \emph {et~al.}(2015)\citenamefont
  {Pierangeli}, \citenamefont {Di~Mei}, \citenamefont {Conti}, \citenamefont
  {Agranat},\ and\ \citenamefont {DelRe}}]{DelRe}%
  \BibitemOpen
  \bibfield  {author} {\bibinfo {author} {\bibfnamefont {D.}~\bibnamefont
  {Pierangeli}}, \bibinfo {author} {\bibfnamefont {F.}~\bibnamefont {Di~Mei}},
  \bibinfo {author} {\bibfnamefont {C.}~\bibnamefont {Conti}}, \bibinfo
  {author} {\bibfnamefont {A.~J.}\ \bibnamefont {Agranat}}, \ and\ \bibinfo
  {author} {\bibfnamefont {E.}~\bibnamefont {DelRe}},\ }\bibfield  {title}
  {\enquote {\bibinfo {title} {Spatial rogue waves in photorefractive
  ferroelectrics},}\ }\href {\doibase 10.1103/PhysRevLett.115.093901}
  {\bibfield  {journal} {\bibinfo  {journal} {Phys. Rev. Lett.}\ }\textbf
  {\bibinfo {volume} {115}},\ \bibinfo {pages} {093901} (\bibinfo {year}
  {2015})}\BibitemShut {NoStop}%
\bibitem [{\citenamefont {Mattheakis}\ \emph {et~al.}(2016)\citenamefont
  {Mattheakis}, \citenamefont {Pitsios}, \citenamefont {Tsironis},\ and\
  \citenamefont {Tzortzakis}}]{Mattheakis}%
  \BibitemOpen
  \bibfield  {author} {\bibinfo {author} {\bibfnamefont {M.}~\bibnamefont
  {Mattheakis}}, \bibinfo {author} {\bibfnamefont {I.J.}\ \bibnamefont
  {Pitsios}}, \bibinfo {author} {\bibfnamefont {G.P.}\ \bibnamefont
  {Tsironis}}, \ and\ \bibinfo {author} {\bibfnamefont {S.}~\bibnamefont
  {Tzortzakis}},\ }\bibfield  {title} {\enquote {\bibinfo {title} {Extreme
  events in complex linear and nonlinear photonic media},}\ }\href {\doibase
  http://doi.org/10.1016/j.chaos.2016.01.008} {\bibfield  {journal} {\bibinfo
  {journal} {Chaos, Solitons and Fractals}\ }\textbf {\bibinfo {volume} {84}},\
  \bibinfo {pages} {73 -- 80} (\bibinfo {year} {2016})}\BibitemShut {NoStop}%
\bibitem [{\citenamefont {Fedele}\ \emph {et~al.}(2016)\citenamefont {Fedele},
  \citenamefont {Brennan}, \citenamefont {Ponce~de Le\'on}, \citenamefont
  {Dudley},\ and\ \citenamefont {Dias}}]{DudleySciRep2016}%
  \BibitemOpen
  \bibfield  {author} {\bibinfo {author} {\bibfnamefont {Francesco}\
  \bibnamefont {Fedele}}, \bibinfo {author} {\bibfnamefont {Joseph}\
  \bibnamefont {Brennan}}, \bibinfo {author} {\bibfnamefont {Sonia}\
  \bibnamefont {Ponce~de Le\'on}}, \bibinfo {author} {\bibfnamefont {John}\
  \bibnamefont {Dudley}}, \ and\ \bibinfo {author} {\bibfnamefont
  {Fr\'ed\'eric}\ \bibnamefont {Dias}},\ }\bibfield  {title} {\enquote
  {\bibinfo {title} {Real world ocean rogue waves explained without the
  modulational instability},}\ }\href {\doibase 10.1038/srep27715} {\bibfield
  {journal} {\bibinfo  {journal} {Scientific Reports}\ }\textbf {\bibinfo
  {volume} {6}},\ \bibinfo {pages} {27715} (\bibinfo {year}
  {2016})}\BibitemShut {NoStop}%
\bibitem [{\citenamefont {Birkholz}\ \emph {et~al.}(2016)\citenamefont
  {Birkholz}, \citenamefont {Br\'ee}, \citenamefont {Veseli\'c}, \citenamefont
  {Demircan},\ and\ \citenamefont {Steinmeyer}}]{Steinmeyer2016}%
  \BibitemOpen
  \bibfield  {author} {\bibinfo {author} {\bibfnamefont {Simon}\ \bibnamefont
  {Birkholz}}, \bibinfo {author} {\bibfnamefont {Carsten}\ \bibnamefont
  {Br\'ee}}, \bibinfo {author} {\bibfnamefont {Ivan}\ \bibnamefont
  {Veseli\'c}}, \bibinfo {author} {\bibfnamefont {Ayhan}\ \bibnamefont
  {Demircan}}, \ and\ \bibinfo {author} {\bibfnamefont {Gunter}\ \bibnamefont
  {Steinmeyer}},\ }\bibfield  {title} {\enquote {\bibinfo {title} {Ocean rogue
  waves and their phase space dynamics in the limit of a linear interference
  model},}\ }\href {\doibase 10.1038/srep35207} {\bibfield  {journal} {\bibinfo
   {journal} {Scientific Reports}\ }\textbf {\bibinfo {volume} {6}},\ \bibinfo
  {pages} {35207} (\bibinfo {year} {2016})}\BibitemShut {NoStop}%
\bibitem [{\citenamefont {Peregrine}\ and\ \citenamefont
  {Smith}(1979)}]{Peregrine1979}%
  \BibitemOpen
  \bibfield  {author} {\bibinfo {author} {\bibfnamefont {D.~H.}\ \bibnamefont
  {Peregrine}}\ and\ \bibinfo {author} {\bibfnamefont {R.}~\bibnamefont
  {Smith}},\ }\bibfield  {title} {\enquote {\bibinfo {title} {Nonlinear effects
  upon waves near caustics},}\ }\href {http://www.jstor.org/stable/75169}
  {\bibfield  {journal} {\bibinfo  {journal} {Philosophical Transactions of the
  Royal Society of London. Series A, Mathematical and Physical Sciences}\
  }\textbf {\bibinfo {volume} {292}},\ \bibinfo {pages} {341--370} (\bibinfo
  {year} {1979})}\BibitemShut {NoStop}%
\bibitem [{\citenamefont {Barkhofen}\ \emph {et~al.}(2013)\citenamefont
  {Barkhofen}, \citenamefont {Metzger}, \citenamefont {Fleischmann},
  \citenamefont {Kuhl},\ and\ \citenamefont {St\"ockmann}}]{Stockmann}%
  \BibitemOpen
  \bibfield  {author} {\bibinfo {author} {\bibfnamefont {S.}~\bibnamefont
  {Barkhofen}}, \bibinfo {author} {\bibfnamefont {J.~J.}\ \bibnamefont
  {Metzger}}, \bibinfo {author} {\bibfnamefont {R.}~\bibnamefont
  {Fleischmann}}, \bibinfo {author} {\bibfnamefont {U.}~\bibnamefont {Kuhl}}, \
  and\ \bibinfo {author} {\bibfnamefont {H.-J.}\ \bibnamefont {St\"ockmann}},\
  }\bibfield  {title} {\enquote {\bibinfo {title} {Experimental observation of
  a fundamental length scale of waves in random media},}\ }\href {\doibase
  10.1103/PhysRevLett.111.183902} {\bibfield  {journal} {\bibinfo  {journal}
  {Phys. Rev. Lett.}\ }\textbf {\bibinfo {volume} {111}},\ \bibinfo {pages}
  {183902} (\bibinfo {year} {2013})}\BibitemShut {NoStop}%
\bibitem [{\citenamefont {Kravtsov}\ and\ \citenamefont
  {Orlov}(1993)}]{Kravtsov}%
  \BibitemOpen
  \bibfield  {author} {\bibinfo {author} {\bibfnamefont {Yu.A.}\ \bibnamefont
  {Kravtsov}}\ and\ \bibinfo {author} {\bibfnamefont {Yu.I.}\ \bibnamefont
  {Orlov}},\ }\href@noop {} {\emph {\bibinfo {title} {Caustics, Catastrophes
  and Wave Fields}}}\ (\bibinfo  {publisher} {Springer},\ \bibinfo {year}
  {1993})\BibitemShut {NoStop}%
\bibitem [{\citenamefont {Mercier}(1962)}]{Mercier}%
  \BibitemOpen
  \bibfield  {author} {\bibinfo {author} {\bibfnamefont {R.~P.}\ \bibnamefont
  {Mercier}},\ }\bibfield  {title} {\enquote {\bibinfo {title} {Diffraction by
  a screen causing large random phase fluctuations},}\ }\href {\doibase
  10.1017/S0305004100036586} {\bibfield  {journal} {\bibinfo  {journal} {Proc.
  Camb. Phil. Soc.}\ }\textbf {\bibinfo {volume} {58}},\ \bibinfo {pages} {382
  -- 400} (\bibinfo {year} {1962})}\BibitemShut {NoStop}%
\bibitem [{\citenamefont {Nye}(1999)}]{Nye}%
  \BibitemOpen
  \bibfield  {author} {\bibinfo {author} {\bibfnamefont {J.~F.}\ \bibnamefont
  {Nye}},\ }\href@noop {} {\emph {\bibinfo {title} {Natural focusing and fine
  structure of light: caustics and wave dislocations}}}\ (\bibinfo  {publisher}
  {Bristol; Philadelphia: Institute of Physics Pub.},\ \bibinfo {year}
  {1999})\BibitemShut {NoStop}%
\bibitem [{\citenamefont {Boyd}(2008)}]{BoydNL}%
  \BibitemOpen
  \bibfield  {author} {\bibinfo {author} {\bibfnamefont {Robert~W.}\
  \bibnamefont {Boyd}},\ }\href@noop {} {\emph {\bibinfo {title} {Nonlinear
  Optics, Chapter 6}}}\ (\bibinfo  {publisher} {Academic Press},\ \bibinfo
  {year} {2008})\BibitemShut {NoStop}%
\bibitem [{\citenamefont {Loudon}(2000)}]{Loudon}%
  \BibitemOpen
  \bibfield  {author} {\bibinfo {author} {\bibfnamefont {Rodney}\ \bibnamefont
  {Loudon}},\ }\href@noop {} {\emph {\bibinfo {title} {The Quantum Theory of
  Light (3rd edition)}}}\ (\bibinfo  {publisher} {Oxford University Press},\
  \bibinfo {year} {2000})\BibitemShut {NoStop}%
\bibitem [{\citenamefont {Steck}(2007)}]{Daniel}%
  \BibitemOpen
  \bibfield  {author} {\bibinfo {author} {\bibfnamefont {Daniel~A.}\
  \bibnamefont {Steck}},\ }\href@noop {} {\emph {\bibinfo {title} {Quantum and
  Atom Optics}}}\ (\bibinfo  {publisher} {http://steck.us/teaching},\ \bibinfo
  {year} {2007})\BibitemShut {NoStop}%
\bibitem [{\citenamefont {McCormick}\ \emph {et~al.}(2004)\citenamefont
  {McCormick}, \citenamefont {Solli}, \citenamefont {Chiao},\ and\
  \citenamefont {Hickmann}}]{McCormick}%
  \BibitemOpen
  \bibfield  {author} {\bibinfo {author} {\bibfnamefont {C.~F.}\ \bibnamefont
  {McCormick}}, \bibinfo {author} {\bibfnamefont {D.~R.}\ \bibnamefont
  {Solli}}, \bibinfo {author} {\bibfnamefont {R.~Y.}\ \bibnamefont {Chiao}}, \
  and\ \bibinfo {author} {\bibfnamefont {J.~M.}\ \bibnamefont {Hickmann}},\
  }\bibfield  {title} {\enquote {\bibinfo {title} {Saturable nonlinear
  refraction in hot atomic vapor},}\ }\href {\doibase
  10.1103/PhysRevA.69.023804} {\bibfield  {journal} {\bibinfo  {journal} {Phys.
  Rev. A}\ }\textbf {\bibinfo {volume} {69}},\ \bibinfo {pages} {023804}
  (\bibinfo {year} {2004})}\BibitemShut {NoStop}%
\bibitem [{\citenamefont {Agrawal}(2013)}]{Agrawal}%
  \BibitemOpen
  \bibfield  {author} {\bibinfo {author} {\bibfnamefont {Govind}\ \bibnamefont
  {Agrawal}},\ }\href@noop {} {\emph {\bibinfo {title} {Nonlinear Fiber Optics
  (Fifth Edition)}}}\ (\bibinfo  {publisher} {Academic Press},\ \bibinfo {year}
  {2013})\BibitemShut {NoStop}%
\bibitem [{\citenamefont {Whitham}(1999)}]{Whitham}%
  \BibitemOpen
  \bibfield  {author} {\bibinfo {author} {\bibfnamefont {G.~B.}\ \bibnamefont
  {Whitham}},\ }\href@noop {} {\emph {\bibinfo {title} {Linear and Nonlinear
  Waves}}}\ (\bibinfo  {publisher} {John Wiley and Sons},\ \bibinfo {year}
  {1999})\BibitemShut {NoStop}%
\bibitem [{\citenamefont {Pitaevskii}\ and\ \citenamefont
  {Stringari}(2003)}]{Pitaevskii}%
  \BibitemOpen
  \bibfield  {author} {\bibinfo {author} {\bibfnamefont {L.}~\bibnamefont
  {Pitaevskii}}\ and\ \bibinfo {author} {\bibfnamefont {S.}~\bibnamefont
  {Stringari}},\ }\href@noop {} {\emph {\bibinfo {title} {Bose-Einstein
  Condensation}}}\ (\bibinfo  {publisher} {Oxford Science Publications},\
  \bibinfo {year} {2003})\BibitemShut {NoStop}%
\bibitem [{\citenamefont {Bludov}\ \emph {et~al.}(2009)\citenamefont {Bludov},
  \citenamefont {Konotop},\ and\ \citenamefont {Akhmediev}}]{Bludov}%
  \BibitemOpen
  \bibfield  {author} {\bibinfo {author} {\bibfnamefont {Yu.~V.}\ \bibnamefont
  {Bludov}}, \bibinfo {author} {\bibfnamefont {V.~V.}\ \bibnamefont {Konotop}},
  \ and\ \bibinfo {author} {\bibfnamefont {N.}~\bibnamefont {Akhmediev}},\
  }\bibfield  {title} {\enquote {\bibinfo {title} {Matter rogue waves},}\
  }\href {\doibase 10.1103/PhysRevA.80.033610} {\bibfield  {journal} {\bibinfo
  {journal} {Phys. Rev. A}\ }\textbf {\bibinfo {volume} {80}},\ \bibinfo
  {pages} {033610} (\bibinfo {year} {2009})}\BibitemShut {NoStop}%
\end{thebibliography}%

\end{document}